\documentclass[reprint,prd,tightenlines,nofootinbib]{revtex4-2}
\usepackage{hyperref}
\hypersetup{colorlinks,allcolors=black}
 \usepackage{graphicx}
 \usepackage{float}
 \usepackage{amsmath}
 \usepackage{amssymb}
 \usepackage[utf8]{inputenc}
 \usepackage{natbib}
 \usepackage{slashed}
 \begin{document}
 
 	\title{The Higgs mechanism from the effective theory of the non-minimal gravity action}
 	\author{Suppanat Supanyo$^{1,2}$}
 	\author{Monsit Tanasittikosol$^{1,2}$}
 	\author{Sikarin Yoo-Kong$^3$}
 	\author{Lunchakorn Tannukij$^4$}
 		\affiliation{$^1$Theoretical and Computational Physics (TCP) Group, Department of Physics, Faculty of Science, King Mongkut's University of Technology Thonburi, Bangkok 10140, Thailand}
 		\affiliation{$^2$Theoretical and Computational Science Centre (TaCS), Faculty of Science, King Mongkut's University of Technology Thonburi, Bangkok 10140, Thailand}
 	\affiliation{$^3$The institute for Fundamental Study (IF), Naresuan University, Phitsanulok 65000, Thailand}
 	\affiliation{$^4$KOSEN-KMITL, King Mongkut's Institute of Technology Ladkrabang, Chalongkrung Road, Ladkrabang, Bangkok 10520, Thailand.}
 		\begin{abstract}
 		 The spontaneous symmetry breaking for the massless scalar field naturally arises from the framework of the effective theory (the non-minimal coupling of gravity to a scalar field). A magic key ingredient is to add the large vacuum energy density,  contributing to the cosmological constant, to the Lagrangian density. 
 		  By applying this modified spontaneous symmetry breaking with the gauge theory (called modified Higgs mechanism), the inflation physics and the electroweak phase transition can be generated from the same framework. However, this comes with the huge price-the large cosmological constant which is known as the dark energy problem.  The possible solution of this issue is also discussed. 
 		  
 		  
 		\end{abstract}
 	\maketitle
\section{\label{sec:intro}Introduction}
It is well known that the Higgs boson plays an essential role in the modern particle physics for giving mass to the elementary particles in the Standard Model \cite{originOfHiggs1, originOfHiggs2, originOfHiggs3, oringPfHiggs4, Electroweak1, Electroweak2, Electroweak3}. This particle also provides the portal, the argument from the naturalness \cite{HeirarchyProblem} and the search for the real shape of the Higgs potential \cite{Maxi-sizing, Probing-quartic}, to the new physics beyond the Standard Model. Moreover, in the context of cosmology, this particle could be a candidate to drive the cosmic inflation \cite{HiggsInflation} and could possibly connect the dark matter to the Standard model particles \cite{HiggsPortal1, HiggsPortal2, HiggsPortal3, HiggsPortal4, HiggsPortal5}. These give us strong evidences that the Higgs particle could play the important role in many areas of modern particle physics.

 One of the most questionable features of the Higgs mechanism is the real shape of the Higgs potential. Up to now, several famous scenarios have been proposed to explain the origin of the electroweak phase transition based on different types of Higgs potential. 
 
 In the Standard model, the Higgs potential is given by
\begin{align}\label{Higgspotential}
    V_\text{Higgs}=-\mu^2 H^\dagger H+\lambda (H^\dagger H)^2,
\end{align}
which is well known as the Landau-Ginzburg potential-type. Here,  $H$ is Higgs field, $\lambda$ is the Higgs self-coupling constant, and $\mu^2$ is the square mass parameter of Higgs field. However, the origin of the mass parameter of the Higgs boson could not be explained by the Higgs mechanism itself. This term is artificially added by hand. Moreover, it artificially requires a mass term with a negative sign. This seems to suggest that there is a missing piece of our understanding about the Standard Model to naturally explain the spontaneous symmetry breaking process of the Higgs boson. With the current experimental technology of collider physics, the search for the real shape of Higgs potential from the Higgs self-coupling sector has not yet come to the conclusion \cite{Observation, Observation2, Observation3, Observation4, Observation5, Observation6, Observation7, Observation8, Observation9, Observation10, Observation11}. 

As a consequence, various ideas have been proposed to  alternatively explain the electroweak phase transition. For example, first, the Coleman-Weinberg mechanism shows that the unusual sign of the mass term could be a result of the quantum correction at one-loop level \cite{ColemanWeinberg} and the electroweak phase transition is triggered by renormalization group running effect \cite{CWRG}. Second, the Tadpole-induced Higgs, the electroweak phase transition is triggered by the Higgs tadpole. However, in this scenario still artificially requires the addition of the positive Higgs mass parameter to the potential. Another possible explanation is that the spontaneous symmetry breaking can be induced from by introducing the dilaton field into the scale-invariant Lagrangian \cite{DilatonModel, DilatonModel2, DilatonModel3}. The mass parameter of Higgs boson in this approach is not required.

Let us give you an overview of the Higgs field coupled non-minimally with gravity. The notable action is given by
\begin{eqnarray}
S=\int d^4x\sqrt{-g}\left\{X-V(\phi)-\frac{M^2+\xi \phi^2}{2}R\right\}\;,\label{action0}
\end{eqnarray}
which has long been studied in the inflationary physics \cite{CiteTheActionOfHiggsInflation, CiteTheActionOfHiggsInflation2, CiteTheActionOfHiggsInflation3, CiteTheActionOfHiggsInflation4, CiteTheActionOfHiggsInflation5, CiteTheActionOfHiggsInflation6, CiteTheActionOfHiggsInflation7} and can be originated from the extension of the Standard Model, involving, the string theory, supersymmetry. Here, $H^\dagger=(0\quad\phi)/\sqrt{2}$ is the Higgs field in the unitary gauge, $X=g^{\mu\nu}\partial_\mu\phi\partial_\nu\phi/2$ is the Higgs kinetic energy, $R$ is the scalar curvature, $M$ is an arbitrary mass scale, $\xi$ is the dimensionless coupling constant, and  $V(\phi)$ is the Higgs potential in the Landua-Ginzberg form \eqref{Higgspotential}.

Traditionally, the action with non-minimal interaction of particle  to gravity is so called the Jordan frame gravity.  Under the conformal transformation, particles couple minimally to gravity through the determinant of metric tensor known as the Einstein frame gravity. In this Einstein frame, the Higgs potential is an effective potential including the effect from the gravity. The interaction $\xi\phi^2 R/2$ can obviously alter the shape of Higgs potential 
\begin{align}
     \tilde{V}(\phi(\tilde{\phi}))=\frac{M_p^4}{(M_p^2+\xi \phi(\tilde{\phi})^2)^2}\left(-\frac{\mu^2}{2}\phi^2(\tilde{\phi})^2+\frac{\lambda}{4} \phi(\tilde{\phi})^4\right),
\end{align}
where $\tilde{\phi}$ is the canonical Higgs field in the Einstein frame. This interaction provides the flatness to the Higgs potential at the large value of the Higgs field.  This characteristic shape of potential in the Einstein frame can be appropriately used to explain the behavior of inflation through scalar field in the slow-roll inflation with the set of parameters $\xi\approx 10^4$, $M\approx M_p$ \cite{HiggsInflation,PowerCounting1,Cutoff1,Cutoff2}, where $M_p= 2.44\times10^{18}$ GeV is a reduced Planck mass. Therefore, the Higgs boson could be interpreted as an inflaton. However, in the case that the square mass parameter of Higgs field vanishes, the self-coupling term still remains in the tree-level potential both Jordan frame and Einstein frame,
\begin{align}
   & V(\phi)=\frac{\lambda}{4} \phi^4,
    \\
    &\tilde{V}(\phi(\tilde{\phi}))=\frac{M_p^4\lambda \phi(\tilde{\phi})^4}{4(M_p^2+\xi \phi(\tilde{\phi})^4)}\approx \frac{\lambda}{4}\tilde{\phi}^4+O(\tilde{\phi}^6),
\end{align}
 respectively. These potentials readily provide the zero tree-level vacuum expectation value (vev) of Higgs field. So the matter and gauge fields receive zero mass after the Higgs field comes to tree-level stability point.  The electroweak phase transition does not happen as we have seen in the experimental observation of the electroweak precision test. Thus, the Colemann-Weiberge mechanism or other mechanisms is required to trigger the electroweak phase transition.
 
 In this work, we shall present that the Higgs mechanism is naturally embedded in the non-minimal gravity action with the perspective of the effective field theory. We first point that the benefit of the interaction $\xi\phi^2 R/2$ is not only to explain the inflation of the universe in the large value of Higgs field in the Einstein frame, but in the small value of Higgs field fluctuating around the electroweak scale, this type of interaction can alternatively trigger the electroweak phase transition without the negative squared mass term. Another key ingredient to pursue our aim is the artificial addition of a constant vacuum energy density ($\rho$) or so called the cosmological constant to the action,
 \begin{align}
S=\int d^4x\sqrt{-g}\left\{-\rho+X-\frac{\lambda}{4}\phi^4-\frac{M^2+\xi\phi^2}{2}R\right\}. \label{action1}
\end{align}
 We simply set  $M\simeq M_p$, and the value of $\xi$ is restricted in the positive region to give a reasonable explanation both the inflationary physics and the particle physics, since the vanishing of mass scale $M$ and coupling constant $\xi$ leads the infinite Higgs mass \cite{induce1,induce2} and too much matter fluctuation for inflation \cite{M0xi0}, respectively.
 

This paper is organized as follows. In Sec-\ref{sec:2}, we study the spontaneous symmetry breaking with the toy-model of the massless scalar field in the action \eqref{action1}. In Sec-\ref{sec:3}, we consider this model of scalar field to be the Higgs field in the unitary gauge to answer the following question: What is the possible value of the parameters $\lambda$, $\rho$, and $\xi$?. Then, in Sec-\ref{sec:4}, we give a concluding summary together with some remarks. 
\section{\label{sec:2}Spontaneous Symmetry breaking of massless scalar field: The Toy model}
 We introduce the conformal transformation of the metric tensor \cite{ConformalTransformation, ConformalTransformation2, ConformalTransformation3, ConformalTransformation4}
\begin{eqnarray}\label{gtransform}
g_{\mu\nu}=\Omega^{-2} \tilde{g}_{\mu\nu}\;,
\end{eqnarray} 
where the parameter with tilde notation refers to the quantity in the Einstein frame. The transformations of the inverse metric tensor and its determinant are
\begin{eqnarray}\label{gtransform2}
g^{\mu\nu}=\Omega^{2} \tilde{g}^{\mu\nu},\quad \sqrt{-g}=\Omega^{-4}\sqrt{-\tilde{g}}.
\end{eqnarray} 
Under these transformations, the kinetic energy of the scalar field and the Ricci scalar curvature become 
\begin{eqnarray}
	&X=g^{\mu\nu}\partial_\mu\phi\partial_\nu\phi=\Omega^2 \tilde{g}^{\mu\nu}\partial_\mu\phi\partial_\nu\phi=\Omega^2 \tilde{X},\label{Xtransform}
	\\
&R=\Omega^{2}\tilde{R}+6 \Omega \tilde{g}^{\mu\nu} \tilde{\nabla}_\mu\tilde{\nabla}_\nu \Omega-12\tilde{g}^{\mu\nu}\tilde{\nabla}_\mu\Omega\tilde{\nabla}_\nu\Omega.\label{Rtransform}
\end{eqnarray}
Using Eq.~\eqref{Rtransform}, the last term in the action \eqref{action1} can be rewritten as
\begin{align}
\nonumber
-&\frac{M_p^2}{2}\int d^4x \sqrt{-g}
\left(1+\frac{\xi\phi^2}{M_p^2}\right)R
\\ \nonumber
=&-\frac{M_p^2}{2}\int d^4x\sqrt{-\tilde{g}} \Omega^{-2}\left(1+\frac{\xi\phi^2}{M_p^2}\right)\tilde{R}
\\\nonumber
&-\frac{M_p^2}{2}\int d^4x\sqrt{-\tilde{g}} \Omega^{-2}\left(1+\frac{\xi\phi^2}{M_p^2}\right)
\\ 
&\times\left(6 \Omega^{-1} \tilde{g}^{\mu\nu} \tilde{\nabla}_\mu\tilde{\nabla}_\nu \Omega-12\tilde{g}^{\mu\nu}\Omega^{-2}\tilde{\nabla}_\mu\Omega\tilde{\nabla}_\nu\Omega\right).\label{ST2thirdterm}
\end{align}
To decouple the non-minimal coupling term of the scalar field and the scalar curvature in the conformal frame, the conformal factor is
\begin{align}\label{Conformalfactor}
\Omega^{2}=1+\frac{\xi\phi^2}{M_p^2}.
\end{align}
Substituting Eq.~\eqref{Conformalfactor} into Eq.~\eqref{ST2thirdterm} and performing integration by parts on  the second order derivative term ($\tilde{g}^{\mu\nu}\nabla_{\mu}\nabla_\nu\Omega$), we obtain
\begin{align}\label{ST2thirdtermfinal}
\nonumber
-&\frac{M_p^2}{2}\int d^4x \sqrt{-g}
\left(1+\frac{\xi\phi^2}{M_p^2}\right)R
\\
=&-\frac{M_p^2}{2}\int d^4x \sqrt{-\tilde{g}}\left(\tilde{R}-\frac{6\xi^2\phi^2}{M_p^4(1+\frac{\xi\phi^2}{M_p^2})^2} \tilde{X}\right)\;.
\end{align}
Next, applying the metric transformations \eqref{gtransform}-\eqref{gtransform2} to the vacuum energy density and the scalar field's kinetic and potential terms, we have
\begin{align}\label{STT2firsttermfinal}
\nonumber\int d^4x \sqrt{-g}&
\left(X-V(\phi)\right)
\\
=&\int d^4x \sqrt{-\tilde{g}}\left(\frac{1}{1+\frac{\xi \phi^2}{M_p^2}}\tilde{X}-\tilde{V}(\phi)\right)\;,
\end{align}
where the potential term is defined as
\begin{align}\label{potentialE}
V(\phi)=\rho+\frac{\lambda}{4}\phi^4\;,
\end{align}
and the transformed potential is
\begin{align}\label{potentialJ}
\tilde{V}(\phi)=\Omega^{-4}\left(\rho+\frac{\lambda}{4}\phi^4\right)=\frac{\rho+\frac{\lambda}{4}\phi^4}{\left(1+\frac{\xi \phi^2}{M_p^2}\right)^2}.
\end{align}

Upon using Eq.~\eqref{ST2thirdtermfinal} and Eq.~\eqref{STT2firsttermfinal}, the full action in the Jordan frame \eqref{action1} can be expressed in the Einstein frame as
\begin{align}\label{ST3}
 S=&\int d^4x \sqrt{-\tilde{g}}
\left(\frac{1}{2}F(\phi)\tilde{X}-\tilde{V}(\phi)-\frac{M_p^2}{2}\tilde{R}\right)\;.
\end{align}
Here the dynamics of the spacetime fluctuation is specified by the metric tensor $\tilde{g}_{\mu\nu}$ and the kinetic energy of scalar field is in the non-canonical form, where
\begin{align}\label{F}
F(\phi)=\frac{M_p^2 \left(M_p^2+\xi  (6 \xi +1) \phi ^2\right)}{\left(M_p^2+\xi  \phi ^2\right){}^2}.
\end{align}
We then redefine the scalar field $\phi$ to obtain the kinetic energy term in the renomalized canonical form  by
\begin{align}\label{redefinefield}
    \frac{d \tilde{\phi}}{d \phi}=\sqrt{F(\phi)}.
\end{align}
The expression for the field redefinition \eqref{redefinefield} in the  small value of $\phi$ ($\xi\phi^2\ll M_p^2$) up to order $\mathcal{O}(M_p^{-2})$ and in the large value of $\phi$ ($\xi\phi^2\gg M_p^2$) are
\begin{align}
    &\phi= \tilde{\phi}\left(1-\frac{\xi\left(6 \xi -1\right) }{6 M_p^2}\tilde{\phi}^2+\mathcal{O}(M_p^{-4})\right)\label{small},
    \\
    &\phi\approx\frac{M_p}{\sqrt{\xi}} \exp\left(\frac{\sqrt{\xi}\tilde{\phi}}{\sqrt{6\xi+1}M_p}\right)\label{large}.
\end{align}
Here, we obtain the perturbative bounds of scalar field in the Einstein frame that is $\tilde{\phi}<\sqrt{6/|\xi(6\xi-1)|}M_p$ for the small value of field, and $\tilde{\phi}>\sqrt{6+(1/\xi)}M_p$ for the large value of field. If the value of $\tilde{\phi}$ is beyond these bounds, the perturbativity of field redefinition therefore breaks down in our approximation.

Substituting Eq.~\eqref{redefinefield} to \eqref{ST3}, we obtain the scalar field with the renormalized canonical kinetic energy coupling minimally with the scalar curvature 
\begin{align}\label{ST4}
 S=&\int d^4x \sqrt{-\tilde{g}}
\left(\frac{1}{2}\partial_\mu\tilde{\phi}\partial^\mu\tilde{\phi}-\tilde{V}(\phi(\tilde{\phi}))-\frac{M_p^2}{2}\tilde{R}\right)\;.
\end{align}
Now, the scalar potential is an effective potential encoded with the interaction from the non-minimal coupling to gravity.

\begin{figure}[H]
 	\centering
 	\includegraphics[width=0.8\linewidth]{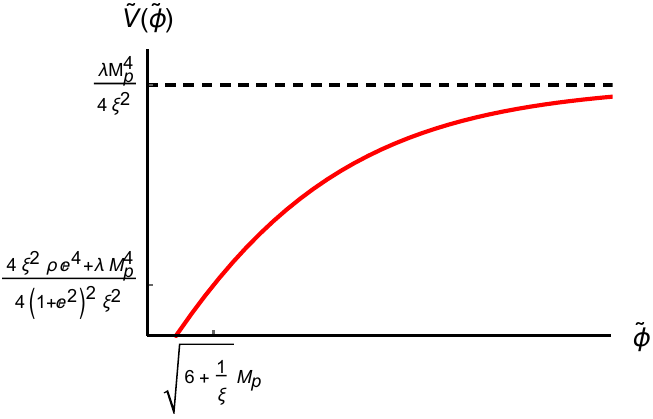}
 	\caption{The sketch of the potential in Eq.~\eqref{V6high}. The dashed line show an asymptotic value of the potential. }
 	\label{fig:PotentialLarge}
 \end{figure}

$\bullet$ The large value of field ($\tilde{\phi}\gg\sqrt{6+(1/\xi)}M_p$):  The potential in the Einstein frame \eqref{potentialJ} is given by substituting Eq.~\eqref{large} to Eq.~\eqref{potentialJ} resulting in
\begin{align}\label{V6high}
    \tilde{V}(\tilde{\phi})=\frac{\rho+\frac{\lambda  M_p^4 }{4 \xi
   ^2}\exp\left(\frac{4 \sqrt{\xi } \tilde{\phi }}{\sqrt{6 \xi +1}}\right) }{\left(1+\exp\left(\frac{2 \sqrt{\xi } \tilde{\phi }}{\sqrt{6 \xi +1}}\right)\right)^2}.
\end{align}
The potential starts to increases from the $\tilde{V}=e^4 \lambda  M_p^4+4 \xi ^2 \rho/4 \left(1+e^2\right)^2 \xi ^2$ at $\tilde{\phi}=\sqrt{6+(1/\xi)}M_p$ to asymptotic constant value,
\begin{align}\label{V6highh}
     \tilde{V}(\tilde{\phi})\approx\frac{\lambda M_p^4}{4\xi^2},
\end{align}
 as shown in Fig.~\ref{fig:PotentialLarge}.
\noindent This behavior is not new in this framework. We can frequently see this potential in the Higgs inflation model  \cite{CiteTheActionOfHiggsInflation, CiteTheActionOfHiggsInflation2, CiteTheActionOfHiggsInflation3, CiteTheActionOfHiggsInflation4, CiteTheActionOfHiggsInflation5, CiteTheActionOfHiggsInflation6, CiteTheActionOfHiggsInflation7}. If this scalar field is interpreted to be Higgs field with  $\lambda=m_h^2/2\upsilon^2$ and $\xi\approx 10^4$, this model turns to be similar model as the Higgs inflation with the same set of slow-roll parameters. Then, it could drive the inflation in the early universe with slow-rolling down from asymptotic value of potential. 

$\bullet$ The small value of field ($\tilde{\phi}\ll \sqrt{6/|\xi(6\xi-1)|}M_p$): Inserting Eq.~\eqref{small} into Eq.~\eqref{potentialJ}, we obtain 

\begin{align}\label{V6}
    \tilde{V}(\tilde{\phi})=\rho-\frac{2 \xi  \rho  }{M_p^2}\tilde{\phi }^2+\frac{\lambda  }{4}\tilde{\phi }^4-\frac{\lambda(1+3\xi)\xi }{3M_p^2}\tilde{\phi }^6+\mathcal{O}(M_p^{-4}),
\end{align}
where $\mathcal{O}(M_p^{-4})$ includes the contribution from $\tilde{\phi}{}^4$ and $\tilde{\phi}{}^6$, and  $\tilde{\phi}{}^8$. Here, we notice that the effective value of ultraviolet cutoff (UV), $\Lambda$, can be specified by flipping the coefficient of the dimension-6 operator and the scalar field  $\tilde{\phi}$ is therefore valid below the cutoff,
\begin{align}\label{cutoff}
    \Lambda\simeq\frac{M_p }{\sqrt{\lambda(1+3\xi)\xi}}.
\end{align}
Moreover, the scalar field obtains the effective square mass term ($m_\phi^2\phi^2/2$) with negative sign as
\begin{align}
    m_\phi^2=-\frac{4\xi\rho}{M_p^2}.
\end{align}
Next, we are going to analyse the potential in two-different situations.

\begin{figure}[H]
 	\centering
 	\includegraphics[width=0.7\linewidth]{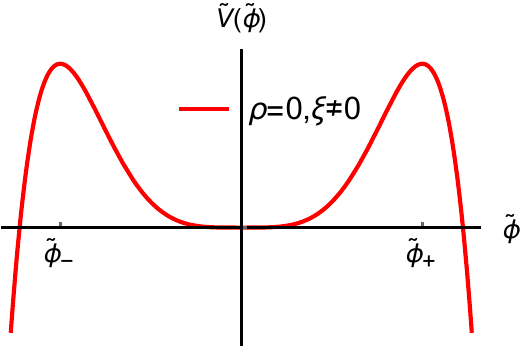}
 	\caption{The sketch of potential in Eq.~\eqref{V6} with $\rho=0$. }
 	\label{fig:V6rho0}
 \end{figure}

Case-1:  $\rho=0$ and $\xi\neq0$, the negative square mass term vanishes. Thus, this scalar field is massless particle with the self-interaction. There are three-extrema points for the potential in Eq.~\eqref{V6},
\begin{align}\label{phiphiplus}
    \phi_0=0,\quad \phi_\pm=\pm\frac{M_p}{\sqrt{2\xi (1+3\xi)}},
\end{align}
 The potential reaches its pseudo maxima at  the $\phi_\pm$ generated from $\tilde{\phi}^6$ operator in $\tilde{V}(\tilde{\phi})$ (The actual highest maximum value of $\tilde{V}(\tilde{\phi})$ can be seen in Eq.~\eqref{V6highh}). On the other hand, the potential attains its minimum value at  $\phi_0=0$, resulting in zero vev ($v$) of $\tilde{\phi}$ \begin{align}
   \phi_0=v=0,
\end{align}
 shown in Fig.~\ref{fig:V6rho0}.
\noindent Obviously, the spontaneous symmetry breaking in tree-level cannot  happen in this case.  After the scalar field rolling down from large scale potential to the stability point at $v=0$, this crucially leads to the zero mass of matter and gauge boson. The electroaweak phase transition cannot be explained by this type of potential. 

\begin{figure}[H]
 	\centering
 	\includegraphics[width=0.7\linewidth]{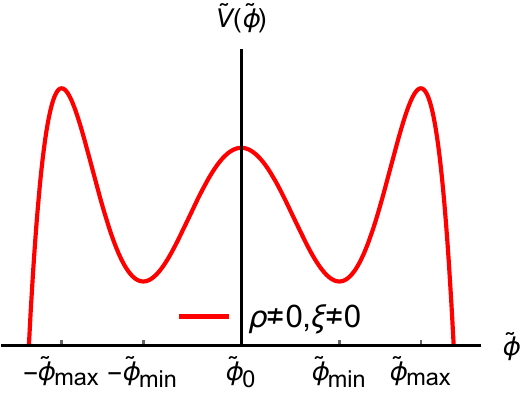}
 	\caption{The sketch of potential in Eq.~\eqref{V6} with $\rho\neq0$}
 	\label{fig:V6rhoneq0}
 \end{figure}

Case-2: of $\rho\neq 0$ and $\xi\neq 0$, the tree-level vacuum energy density, $\rho$, and the coupling constant from non-minimal interaction with gravity, $\xi$, influence the ground state of scalar field in the Einstein frame, shown in Fig.~\ref{fig:V6rhoneq0}.
 \noindent The $\tilde{\phi}_\text{max}$ is the pseudo maximum point of potential in the same analogy with $\tilde{\phi}_\pm$ in Eq.~\eqref{phiphiplus}. 
In this case, the vev of $\tilde{\phi}$ is non-zero, obtained at $\tilde{\phi}_\text{min}$. It is given by 
\begin{align}\label{vev0}
    v^2=\frac{M_p^2}{2\xi(1+3\xi)}-\frac{M_p^2}{2\xi(1+3\xi)}\sqrt{1-\frac{32\xi^2(1+3\xi)\rho}{\lambda M_p^4}}.
\end{align}
This then triggers the spontaneous symmetry breaking. The value of $v$ is real, if $\rho<\lambda M_p^4/32\xi^2(1+3\xi)$ which gives us the upper limit of $\rho$. When the value of $\rho$ tends to $\lambda M_p^4/32\xi^2(1+3\xi)$, the vev of scalar field becomes
\begin{align}
    v^2\approx \frac{M_p^2}{2\xi(1+3\xi)}.
\end{align}
If this model is set to be inflation model by choosing $\xi\approx 10^4$, the interpretation of this scalar field to be Higgs field is ruled out. The square vev of the scalar field is around ten orders of magnitude larger than the square of Electroweak vev. On the other hand, if this scalar field is set to be Higgs boson with $\upsilon\simeq 10^2$ GeV, the coupling constant $\xi$ is around sixteen orders of magnitude larger than the Higgs inflation model.  Hence, the large value of $\rho\approx \lambda M_p^4/32\xi^2(1+3\xi)$ gives a conflict explanation between the inflationary physics and particle physics.

Furthermore, if we consider $\rho\lll\lambda M_p^4/32\xi^2(1+3\xi)$, we obtain
\begin{align}\label{vev}
v^2\approx&\frac{M_p^2}{2\xi(1+3\xi)}-\frac{M_p^2}{2\xi(1+3\xi)}+\frac{8\xi\rho}{\lambda M_p^2}=\frac{8\xi\rho}{\lambda M_p^2}.
\end{align}
In this limit, the vev can be much smaller than below the Planck scale although the value of $\xi$ is approximately ten to the fourth. The value of $\xi\rho/\lambda$  is required to be four orders of magnitude larger than $M_p^2$ to generate the electroweak vev. This regime of $\rho$ could possibly reconcile both the inflationary physics and the particle physics. This then naturally leads to the spontaneous symmetry breaking of the scalar field with the hierarchy between its vev and the Planck scale. This behavior commonly appears in the Electroweak theory.

  In the next section, we then interpret the Einstein frame as a physical frame, and the $\tilde{\phi}$ is the fundamental field for unitary Higgs field with the renormalized canonical kinetic energy in the Electroweak theory. We apply the modified Mexican-hat potential in Eq.~\eqref{V6}, see Fig.~\ref{fig:V6rhoneq0}, to fit the parameters $\lambda,\rho,\xi$ with Fermi-coupling constant, Higgs mass, and cosmological constant in tree-level.

\section{\label{sec:3}Higgs Boson}
Let us start with promoting the scalar field $\phi$ in Eq.~\eqref{action1} as the Higgs field (Jordan frame) in the unitary gauge 
\begin{align}\label{Higgsdoublet}
    H(x)=\frac{1}{\sqrt{2}}\begin{pmatrix}
    0\\
    \phi(x)
    \end{pmatrix},
\end{align}
and adding the gauge field  and leptonic field to the action~\eqref{action1}. We have
\begin{align}\label{Weakaction}
    S=&\int d^4x\sqrt{-g}\left\{ \mathcal{L}_\text{Higgs} +\mathcal{L}_\text{Gauge}+\mathcal{L}_\text{Lepton}+ \mathcal{L}_\text{Gravity}\right\},
\end{align}
where 
\begin{align}
    \mathcal{L}_\text{Higgs}=&g^{\mu\nu}(D_\mu H)^\dagger D_\nu H-\rho -\lambda (H^\dagger H)^2,\label{LhiggsJ}
    \\
    \mathcal{L}_\text{Gauge}=&-\frac{1}{2}g^{\mu\nu}g^{\alpha\beta}W^+_{\mu\nu}W^-_{\alpha\beta}-\frac{1}{4}g^{\mu\nu}g^{\alpha\beta}Z_{\mu\nu}Z_{\alpha\beta}\nonumber
     \\
     &-\frac{1}{4}g^{\mu\nu}g^{\alpha\beta}A_{\mu\nu}A_{\alpha\beta}\label{LgaugeJ},
    \\
    \mathcal{L}_\text{Lepton}=& i\overline{L} \slashed{D} L+i\overline{R} \left(\slashed{\partial}+\frac{ig'}{2}\slashed{B}\right) R\nonumber
    \\
    &-\lambda_f \left(\overline{R}H^\dagger L+\overline{L} H R\right),\label{LleptonJ}
 \\
    \mathcal{L}_\text{Gravity}=&-\left(\frac{M_p^2}{2}+\xi H^\dagger H\right)R.\label{LgravityJ}
\end{align}
 Here, $D_\mu$ in Eq.~\eqref{LhiggsJ} is the covariant derivative of $SU(2)\times U(1)$ symmetry given by
\begin{align}\label{covariantderivative}
    D_\mu=\partial_\mu-\frac{ig}{\sqrt{2}}(\sigma^+W_\mu^++\sigma^-W_\mu^-)-\frac{ig}{2}\sigma_3 W_\mu^3+\frac{i g'}{2}B_\mu,
\end{align}
where $g$ and $g'$ are the gauge coupling constant, $\sigma_i$ are Pauli matrices, $W_{\mu}^i$ are $SU(2)$ gauge field, $B_\mu$ is $U(1)$ gauge field. The $W_\mu^3$ and $B_\mu$ can be reorganized in terms of $Z$-field ($Z_\mu$) and photon-field ($A_\mu$) as
 \begin{align}
     W_\mu^3=&Z_\mu \cos(\theta_W)-A_\mu \sin(\theta_W),
     \\
     B_\mu=&Z_\mu \sin(\theta_W)+A_\mu\cos(\theta_W).
 \end{align}
 The $W_{\mu\nu}$, $Z_{\mu\nu}$ and $A_{\mu\nu}$ in Eq.~\eqref{LgaugeJ} are the field strength tensor of $W$, $Z$ boson, and photon, respectively. The $L$ and $R$ in Eq.~\eqref{LleptonJ} are left-right handed spinor doublet field of SU(2) symmetry given by
 \begin{align}
     L=\begin{pmatrix}
    \nu_L
    \\
    \psi_L
    \end{pmatrix},\quad R=\psi_R,
 \end{align}
 which $\nu$ and $\psi$ are generic field of neutrino and lepton, respectively. Lastly, $\lambda_f$ is Yukawa coupling constant.
 
Substituting Eq.~\eqref{Higgsdoublet} into Eq.~\eqref{Weakaction} and then performing the conformal transformation Eqs.~\eqref{gtransform}-\eqref{gtransform2} together with the field redefinition \eqref{small} of unitary Higgs field in Jordan frame, the action~\eqref{Weakaction} in the small value of Higgs field is reorganized as
\begin{align}
    S=&\int d^4x\sqrt{-g'}\nonumber
    \\
    &\times( \mathcal{L}'_\text{Higgs} +\mathcal{L}'_\text{Gauge}+\mathcal{L}'_\text{Lepton}+ \mathcal{L}'_\text{Gravity}+...),
\end{align}
where 
\begin{align}
    \mathcal{L}'_\text{Higgs}\approx&\frac{1}{2}\partial_\mu \tilde{\phi}\partial^\mu \tilde{\phi}-\rho+\frac{2\xi\rho}{M_p^2}\tilde{\phi}^2-\frac{\lambda}{4}\tilde{\phi}^4+\frac{\lambda(1+3\xi)\xi}{3M_p^2}\tilde{\phi}^6\label{LHiggsE},
    \\
     \mathcal{L}'_\text{Gauge}=&-\frac{1}{2}\tilde{g}^{\mu\nu} \tilde{g}^{\alpha\beta}W^+_{\mu\nu}W^-_{\alpha\beta}+\frac{g^2\phi(\tilde{\phi}^2)}{4\Omega^2}\tilde{g}^{\mu\nu}W_\mu^+W_\nu^-\nonumber
     \\
     &-\frac{1}{4}\tilde{g}^{\mu\nu} \tilde{g}^{\alpha\beta}Z_{\mu\nu}Z_{\alpha\beta}+\frac{g^2\phi(\tilde{\phi}^2)}{8\Omega^2\cos^2(\theta_W)}\tilde{g}^{\mu\nu}Z_\mu Z_\nu\nonumber
     \\
     &-\frac{1}{4}\tilde{g}^{\mu\nu} \tilde{g}^{\alpha\beta}A_{\mu\nu}A_{\alpha\beta}\label{LgaugeE},
    \\
    \mathcal{L}'_\text{Lepton}=&\frac{1}{\Omega^3}\left(i\overline{\psi}\slashed{\partial} \psi+i\overline{\nu}_{L}\slashed{\partial} \nu_{L}\right)-\frac{\lambda_f}{\sqrt{2}\Omega^4}\phi(\tilde{\phi}) \overline{\psi}\psi\nonumber
    \\
    &-\frac{g}{\sqrt{2}\Omega^{3}}(\overline{\nu}_L \slashed{W}^+ \psi_L+\overline{\psi}_L\slashed{W}^-\nu_L)-\frac{e}{\Omega^3}\overline{\psi}\slashed{A}\psi\nonumber
    \\
    &-\frac{1}{2\Omega^3\sqrt{g^2+g'^2}}(g^2[\overline{\nu}_L\slashed{Z}\nu_L-\overline{\psi}_L\slashed{Z}\psi_L]\nonumber
    \\
    &+g'^2[\overline{\nu}_L\slashed{Z}\nu_L+\overline{\psi}_L\slashed{Z}\psi_L+2\overline{\psi}_R\slashed{Z}\psi_R]),
    \\
    \mathcal{L}_\text{Gravity}=&-\frac{M_p^2}{2} \tilde{R},
\end{align}
and we have defined $\psi=\psi_L+\psi_R$. After the transformations, the $\tilde{\phi}$ is interpreted as the canonical Higgs field in unitary gauge with canonical kinetic energy. The kinetic energy terms of gauge boson are invariant therefore we do not need  re-scaled gauge fields to reorganize the kinetic energy in the canonical form. The only change for the gauge bosons is quadratic term coupling to the Higgs field. In the case of fermionic field, the kinetic energy is not the canonical form. We then require the field redefinition,
\begin{align}
    \tilde{\psi}\to\Omega^{3/2}\psi, \quad\nu\to\Omega^{3/2}\nu,
\end{align}
to obtain the renormalized canonical kinetic energy. We have
\begin{align}
    \mathcal{L}'_\text{Lepton}=&i\overline{\psi}\slashed{\partial} \psi+i\overline{\nu}_{L}\slashed{\partial} \nu_{L}-\frac{\lambda_f}{\Omega}\phi(\tilde{\phi}) \overline{\psi}\psi+\frac{3}{2\Omega}i\overline{\psi}\psi  \slashed{\partial}\Omega\nonumber
     \\
    &-\frac{g}{\sqrt{2}}(\overline{\nu}_L \slashed{W}^+ \psi_L+\overline{\psi}_L\slashed{W}^-\nu_L)-e\overline{\psi}\slashed{A}\psi\nonumber
    \\
    &-\frac{1}{2\sqrt{g^2+g'^2}}(g^2[\overline{\nu}_L\slashed{Z}\nu_L-\overline{\psi}_L\slashed{Z}\psi_L]\nonumber
    \\
    &+g'^2[\overline{\nu}_L\slashed{Z}\nu_L+\overline{\psi}_L\slashed{Z}\psi_L+2\overline{\psi}_R\slashed{Z}\psi_R])\label{LleptonE}.
\end{align}
 Now, what we can see is that the fermion-gauge interaction is invariant, while the Yukawa coupling is modified by the conformal factor.
\subsection{\label{sec:3-1}Tree-level fitting the vev to the Fermi coupling constant}
We show the relative deviation of the vev in our model from the Standard Model prediction. 

The vev, $v$, is traditionally defined by matching the electroweak theory to the four-fermion theory in the low energy limit of muon decay channel. We briefly provide the calculation as follows. The coupling constant in the four-fermion theory, so called Fermi coupling constant ($G_F\approx1.166\times10^{-5}$ GeV$^{-2}$ \cite{EWtest}), can be obtained by integrating out the field of W-boson from the weak interaction with lepton, and could be written in terms of the mass of $W^\pm$ boson and gauge coupling constant ($g$)  as 
\begin{align}\label{FermiCoupling}
\frac{g^2}{8M_W^2}=\frac{G_F}{\sqrt{2}},
\end{align}
where $M_W$ is the mass of $W^\pm$ bosons defined as
\begin{align}
\frac{g^2}{4}\phi^2W^{+}_\mu W_{-}^\mu=\frac{g^2v_\text{SM}^2}{4}W^{+}_\mu W_{-}^\mu+\mathcal{O}(h),
\end{align}
after spontaneous symmetry breaking.
The mass of $W^\pm$ bosons can be written in  terms of Higgs vev as 
\begin{align}\label{SMWmass}
M_W=\frac{gv_{\text{SM}}}{2}.
\end{align}
Substituting Eq.~\eqref{SMWmass} to Eq.~\eqref{FermiCoupling}, the  vev can be obtained as follows
\begin{align}\label{nuSMobservation}
v_{\text{SM}}=1/\sqrt{\sqrt{2} G_F}\approx 246.3\text{GeV}.
\end{align}

In our case, the interaction of lepton and W-boson is not modified from the four-fermion theory, the fourth term in Eq.~\eqref{LleptonE}. Therefore, the $G_F$ is not altered from  Eq.~\eqref{FermiCoupling}. However, the mass square of $W$ boson in terms of $g$ and vev is modified by the conformal factor. Expanding $\tilde{\phi}$ around vev in Eq.~~\eqref{LgaugeE}, we obtain
\begin{align}\label{SMWmassE}
    M_W=\frac{g}{2}\frac{\phi(v)}{\Omega(\phi(v))}.
\end{align}
 Substituting  $M_W$ from Eq.~\eqref{SMWmassE} into Eq.~\eqref{FermiCoupling}, we obtain
\begin{align}\label{above}
    \frac{G_F}{\sqrt{2}}=\frac{\Omega^2(\phi(v))}{2\phi^2(v)}.
\end{align}
Then, if we parametrize $G_F$ in Eq.~\eqref{above} with Eq.~\eqref{nuSMobservation}, we obtain the relation of modified Higgs vev from this model and the vev from the electroweak theory as
\begin{align}\label{vevrelation}
    v^2_\text{SM}=\frac{\phi(v)^2}{\Omega(\phi(v))^2}.
\end{align}
By expanding $\tilde{\phi}$ around vev in the second term of Eq.~\eqref{LgaugeE} and Eq.~\eqref{LleptonE}, we obtain 
\begin{align}
    & m_\psi=\lambda_f \left(\frac{\phi(v)}{\Omega(\phi(v))}\right)= \lambda_f v_\text{SM},\label{massf}
     \\
    &  M_W=\frac{g}{2}\left(\frac{\phi(v)}{\Omega(\phi(v))}\right)=\frac{g v_\text{SM}}{2},\label{massW}
\end{align}
which $m_\psi$ is lepton mass. Interestingly, the mass of lepton and gauge boson remains the same with the Standard model prediction even though the vev is modified
\begin{align}\label{vevsmrelation}
v ^2=v _{\text{SM}}^2\left(1+\frac{2\xi\left( 1+3\xi  \right) v_{\text{SM}}^2}{3 M_p^2}+\mathcal{O}(M_p^{-4})\right),
\end{align}
where we have substituted Eq.~\eqref{small} to Eq.~\eqref{vevrelation} and performed the series expansion up to the order $M_p^{-2}$.
Now, the vev of Higgs field cannot be identified in the same ananoly with the electroweak theory since the $\upsilon$ depend on the parameter $\xi$. 


 The convergent condition of Eq.~\eqref{vevsmrelation} gives us the upper bound of $\xi$: $\xi\ll M_p/\sqrt{2}v_{\text{SM}}$ therefore the value of $\xi$ in our model could be in the regime $0<\xi\lesssim M_p/\sqrt{2}v_{\text{SM}}$. This bound of $\xi$ includes the case of inflationary physics at large scale of Higgs field with $\xi\approx 10^4$. The square vev of Higgs field in this model is modified from the square vev in the standard electroweak theory at the order $v^2\sim(1+10^{-24})v_{\text{SM}}^2$. If the value of $\xi$ approaches its upper bound, the value of the square vev will be approximately doubled the value of the square electroweak vev: $v^2\simeq 2v_\text{SM}^2$, with the unchanged of the value of Fermi coupling in Eq.~\eqref{above} and particle mass in Eq.~\eqref{massf} and Eq.~\eqref{massW}.
\subsection{Matching \texorpdfstring{$\rho$}{rho}  and \texorpdfstring{$\lambda$}{lambda}  to Higgs mass and Higgs vev}
We consider the expansion of the Higgs field around the vev,
\begin{align}\label{phipertube}
    \tilde{\phi}(x)=v+h(x),
\end{align}
where $\upsilon$ is defined in Eq.~\eqref{vev0} and $h(x)$ is fluctuation around the vev. Substituting Eq.~\eqref{phipertube} into Eq.~\eqref{V6}, the potential of Higgs field becomes 
\begin{align}\label{V56}
 &\tilde{V}(\phi(v+h))= \frac{1}{2(1+3\xi)}\left((1+6\xi)\rho+\frac{\lambda M_p^4}{48\xi^2(1+3\xi)}\right)\nonumber
    \\
  & +\frac{1}{3(1+3\xi)}\left(\rho-\frac{\lambda M_p^4}{32\xi^2(1+3\xi)}\right)\sqrt{1-\frac{32 \xi ^2 (3 \xi +1) \rho }{\lambda  M_p^4}}\nonumber
   \\
  & +\left(\frac{8 \xi \rho }{ 
   M_p^2}+\frac{ \lambda  M_p^2}{4 \xi  (3 \xi
   +1)}\left[\sqrt{1-\frac{32 \xi ^2 (3 \xi +1) \rho }{\lambda  M_p^4}}-1\right]\right)h^2\nonumber
   \\
   &+\mathcal{O}(h^3).
\end{align}
Here, we consider the expansion in Eq.~\eqref{V56} up to the order $\mathcal{O}(h^2)$, since the mass term comes with $h^2$.
Therefore, the square Higgs mass, denoted by $M_h^2$, is
\begin{align}\label{mass0}
    M_h^2=\frac{16 \xi \rho }{ 
   M_p^2}+\frac{ \lambda  M_p^2}{2 \xi  (3 \xi
   +1)}\left[\sqrt{1-\frac{32 \xi ^2 (3 \xi +1) \rho }{\lambda  M_p^4}}-1\right],
\end{align}
where $M_h\approx 125$ GeV is Higgs mass \cite{HiggsMass125,HiggsMass1252}.
In the limit $\rho\lll\lambda M_p^4/32\xi^2(1+3\xi)$, which gives a consistent explanation for both the inflationary physics and the electroweak theory, we find that the large term proportional to square Planck mass in Eq.~\eqref{mass0} is cancelled out and the Higgs mass is proportional to $M_p^{-2}$ as
\begin{align}\label{mass}
    M_h^2\approx&\frac{ \lambda  M_p^2}{2 \xi  (3 \xi
   +1)}-\frac{ \lambda  M_p^2}{2 \xi  (3 \xi
   +1)}+\frac{16 \xi  \rho }{M_p^2}-\frac{8 \xi  \rho }{M_p^2}
   =\frac{8\xi\rho}{M_p^2}.
\end{align}
Then, from the relation between Eq.~\eqref{vev} and Eq.~\eqref{mass0}, we can solve the parameters $\lambda$ and $\rho$ in terms of $M_h$ and $v$ as follows
\begin{align}
     &\lambda =\frac{M_h^2 M_p^2}{2 v^2 (M_p^2-4 \xi  (3 \xi +1) v^2)}\label{lambda0},
    \\
    &\rho =\frac{M_h^2 M_p^2 \left(M_p^2-2 \xi  (3 \xi +1) v^2\right)}{8 \xi  \left(M_p^2-4 \xi  (3 \xi
   +1) v^2\right)}\label{rho0}.
\end{align}
Substituting Eq.~\eqref{vevsmrelation} into Eq.~\eqref{lambda0} and Eq.~\eqref{rho0}, we obtain the parameters $\lambda$ and $\rho$ in terms of the observable parameters $M_h$ and $v_\text{SM}$ as  
\begin{align}
    \lambda =&\frac{M_h^2}{2 v_{\text{SM}}^2}+\frac{5 \xi  (3 \xi +1) M_h^2}{3
   M_p^2}+\mathcal{O}(M_p^{-4})\label{lambdasm},
    \\
    \rho =&M_p^2\left(\frac{M_h^2 }{8 \xi }+\frac{(3 \xi +1) M_h^2 v_{\text{SM}}^2}{4M_p^2}+\mathcal{O}(M_p^{-4}) \right)\label{rhosm}.
\end{align}
Now, employing the bound of paramter $\xi$,  $0<\xi<M_p/\sqrt{2}v_\text{SM}$, we find that the value of $\lambda$ can possibly reach to six times of the $\lambda_{\rm SM}$, predicted from the standard electroweak theory: $\lambda\approx 6\lambda_\text{SM}$.By substituting the maximum value of $\xi$ to Eq.~\eqref{rhosm}, we can obtain the minimum value of $\rho\gtrsim M_h^2 M_p v_{\text{SM}}/4\sqrt{2}\approx 10^{24}$ GeV$^4$, in order to provide the electroweak phase transition. 
Next, if we are working with the inflaton model, $\xi\approx 10^4$, the Higgs self-coupling constant is modified as $\lambda=(1+10^{-23})\lambda_\text{SM}$ , and the required value of constant vacuum energy density is $\rho\approx 10^{36}$ GeV$^4$.  All possible values of the $\xi$ in the positive regime give to a huge contribution to the vacuum energy density. This then leads to the cosmological constant problem.
\section{\label{ccproblem}Discussion on the Cosmological constant problem}
\noindent The cosmological constant ($\Lambda$) is defined as
\begin{align}
    \Lambda=8\pi G\rho_\text{vacuum},
\end{align}
where $G$ is the gravitational constant and $\rho_\text{vacuum}\approx 2.5\times 10^{-47}$ GeV${}^4$ is the observable value of vacuum energy density. The role of this cosmological constant has been famously used to explain the accelerating expansion of the universe, known in the name of ``dark energy''. However, in our scenario, the role of vacuum energy density $\rho$ is not used to explain the dark energy, but, rather, is used to trigger the symmetry breaking. Now, we consider the constant term, denoted by $\tilde{\rho}$, in Eq.~\eqref{V56},
\begin{align}\label{rhotilde0}
  &\tilde{\rho}=\frac{1}{2(1+3\xi)}\left((1+6\xi)\rho+\frac{\lambda M_p^4}{48\xi^2(1+3\xi)}\right)\nonumber
    \\
  & +\frac{1}{3(1+3\xi)}\left(\rho-\frac{\lambda M_p^4}{32\xi^2(1+3\xi)}\right)\sqrt{1-\frac{32 \xi ^2 (3 \xi +1) \rho }{\lambda  M_p^4}},
\end{align}
which will be seen later contributing to $\rho_{\rm vacuum}$.
Substituting $\lambda$, $\rho$ from Eq.~\eqref{lambdasm}, Eq.~\eqref{rhosm} into Eq.~\eqref{rhotilde0}, we obtain 
\begin{align}\label{rhotildesm}
    \tilde{\rho}=&\frac{M_h^2 M_p^2}{8 \xi }+\frac{1}{8} (6 \xi +1) M_h^2 v_{\text{SM}}^2+O(M_p^{-2}).
\end{align}
Obviously, the $\tilde{\rho}$ is dominated by $M_h^2M_p^2/8\xi$ GeV$^4$. If we set $\tilde{\rho}=\rho_\text{vacuum}$, the value of $\xi$ can be computed from the first term in Eq.~\eqref{rhotildesm}
 \begin{align}
     \xi\simeq \frac{M_h^2 M_p^2}{\rho_\text{vacuum}}\approx 10^{86},
 \end{align}
 which turns out to be a huge number and is not in the range of the convergent condition: $0<\xi< M_p/\sqrt{2}v_\text{SM}$. This suggests that the dimensionless coupling constant $\xi$ could not be specified by the cosmological constant. The big elephant in our scenario seems to be the dark energy problem.  Therefore, we require the accurate cancellation of tree-level vacuum energy density. 
 
 To modify the vacuum energy density so that it agree with the observation, we expect the remaining of $\tilde{\rho}$ in Eq.~\eqref{rhotildesm} to be at the order $O(M_p^{-4})$, which could be parametrically written in terms of
 \begin{align}\label{rhoexpect}
     \tilde{\rho}\sim\frac{M_h^2 v_\text{SM}^6}{M_p^4}\approx 10^{-56}(\text{GeV})^4.
 \end{align}
 We propose one possible way to fix this problem. The kinetic energy of the Higgs field in the Jordan frame may be given in the non-standard form as
 \begin{align}\label{kenew}
     \Lambda_k^4\left(1+\frac{\xi \phi^2}{M_p^2}\right)^2\sqrt{1+\frac{\partial_\mu\phi\partial^\mu \phi}{\left(1+\frac{\xi \phi^2}{M_p^2}\right)^2\Lambda_k^4}}
 \end{align}
 where $\Lambda_k$ is constant with energy-dimension $E^1$. By assuming that $\partial_\mu\phi\partial^\mu\phi/2\ll \Lambda^4_k$, the term $(\partial_\mu\phi\partial^\mu\phi/2\Lambda_k^4)^2$ appearing from the series expansion can be ignored in our calculation.  After the conformal transformation, the lowest order expansion of the kinetic energy \eqref{kenew} does not change from Eq.~\eqref{ST3} so the field redefinition in Eq.~\eqref{redefinefield} is also in the same form. This then results in the shape of the effective potential $\tilde{V}$ in Eq.~\eqref{ST4} not being  spoiled. The value of $\xi\approx 1.76\times10^4$ and the prediction of the parameters $\lambda$ and $\rho$ from Eq.~\eqref{lambdasm}-\eqref{rhosm} are approximately applicable. In the Einstein frame, we fine that the value of the tree-level vacuum energy is 
 \begin{align}
     \tilde{\rho}\to\tilde{\rho}-\Lambda^4_k.
 \end{align}
 To obtain the compatibility between the theoretical prediction and the observed value of the cosmological constant, we have to parametrize the $\Lambda_k^4$ in the following form 
 \begin{align}
     \Lambda_k^4=&\frac{M_h^2 M_p^2}{8 \xi }+\frac{1}{8} (6 \xi +1) M_h^2 v_{\text{SM}}^2+\frac{\xi  (6
   \xi +1) M_h^2 v_{\text{SM}}^4}{8 M_p^2}\nonumber
   \\
   &+\frac{\xi ^2 (18 \xi -17) M_h^2 v_{\text{SM}}^6}{24 M_p^4}.
 \end{align}
  By this definition, the large-value contributions to the cosmological constant in tree-level are cancelled and the remaining of the tree-level cosmological constant  can be compatible with the observation,
 \begin{align}\label{eqx}
    \tilde{\rho}-\Lambda^4_k=\frac{5M_h^2 v_\text{SM}^6\xi^2}{6 M_p^4}+\mathcal{O}(M_p^{-6})\approx 2.5\times10^{-47} \text{(GeV)}^4
 \end{align}
where $\xi\approx1.76\times 10^4$, $M_h=125$ GeV, $v_\text{SM}=246$ GeV. Then, we consider the zero-point energy ($\rho_\text{zero}$) from whole particles in the Standard Model contributing to the vacuum energy density. We apply the on-shell subtraction scheme in our calculation. Here, overall parameters in the action are the physical quantity and the overall loop-corrections can be absorbed into the added counterterms.  Now, the  UV-divergence from $\rho_\text{vacuum}$ can be cancelled out by taking the counterterm $\delta\rho$ into the account,
\begin{align}\label{rhovaccc}
    \rho_{\text{vacuum}}&=\left(\frac{5M_h^2 v_\text{SM}^6\xi^2}{6 M_p^4}+\mathcal{O}(M_p^{-6})\right)+\rho_\text{zero}+\delta \rho.
\end{align}
Under the on-shell condition $\delta\rho=-\rho_\text{zero}$, the remainder in the cosmological constant is
\begin{align}
    \rho_\text{vacuum}\approx\frac{5M_h^2 v_\text{SM}^6 \xi^2}{6M_p^4}\approx 2.5\times 10^{-47} (\text{GeV})^4.
\end{align}
Finally,  we can obtain the compatibility between the $\rho_{\text{vacuum}}$ from the model and the observation of the cosmological constant.
\section{\label{sec:4} Conclusion}
In our scenario, the original theory in Eq.~\eqref{action1} is restricted in the Jordan frame. The spontaneous symmetry breaking and electroweak phase transition are promoted as the effective theory from the gravity.  This frame work might be an alternatively possible way to unify the electroweak phase transition and inflation. Here, a huge cosmological constant from the prediction of the model leads to a major problem. However, this problem could be  solved in the on-shell subtraction scheme when the kinetic energy of the Higgs field is in the non-standard form \eqref{kenew} with the cutoff $\Lambda_k\sim 10^9$ GeV as we have shown in the discussion. We finally note that the presented scheme on modifying kinetic energy is just one of many possible ways to match with the observable cosmological constant.

\begin{acknowledgements}
We acknowledge the support from the Petchra Prajomklao Ph.D. Research Scholarship from King Mongkut’s University of Technology Thonburi (KMUTT). We also would like to thank Associate Professor Dr. Daris Samart and Associate Professor Dr. Chakrit Pongkitivanichkul for valuable discussions.
\end{acknowledgements}
\bibliographystyle{apsrev4-1}
\bibliography{mybib.bib}
 \end{document}